# Elastic and thermodynamic properties of the shape-memory alloy AuZn

T. W. Darling, F. Chu, A. Migliori, D. J. Thoma, M. Lopez

[1]Materials Science and Technology Division, Los Alamos National Laboratory, Los Alamos NM 87545

J. C. Lashley[1,2], B. E. Lang, J. Boerio-Goates, and B. F. Woodfield

[2]Department of Chemistry and Biochemistry, Brigham Young University, Provo UT 84602

The current work reports on the elastic shear moduli, internal friction, and the specific heat of the B2 cubic ordered alloy AuZn as a function of temperature. Measurements were made on single-crystal and polycrystal samples using Resonant Ultrasound Spectroscopy (RUS), semi-adiabatic calorimetry, and stress-strain measurements. Our results confirm that this alloy exhibits the shape-memory effect and a phase transition at 64.75 K that appears to be continuous (second-order) from the specific heat data. It is argued that the combination of equiatomic composition and a low transformation temperature constrain the chemical potential and its derivatives to exhibit behavior that lies at the borderline between that of a first-order (discontinuous) and a continuous phase transition. The acoustic dissipation does not peak at the transition temperature as expected, but shows a maximum well into the low-temperature phase. The value of $Q_D$ 219 K, obtained from the low-temperature $C_p$ data is in favorable agreement with that determined from the acoustic data above the transition, $Q_D = 207$ K.

# I. INTRODUCTION

The engineering impact of shape-memory alloys emanate from the ability of the material to generate force with thermal excursions through a phase transformation.[1] From a simple perspective, induced strain can be recovered from an aligned, lower symmetry phase upon transformation to a higher symmetry phase with increasing temperature. Consequently, the high-temperature shape can be recovered with a structural reversion during a thermal cycle.[2] The phase transformation associated with the shape-memory effect (SME) is the thermoelastic martensite. A martensitic phase transformation is a first-order, displacive reaction with strict rules that determine relative crystallographic orientations between the parent (high-temperature) and product (low-temperature) phases, resulting in twin relationships between variants in the product phase.[2]

The thermoelastic martensitic transformation is a nucleation and growth reaction that proceeds by a shear-like mechanism, and therefore interacts with applied stress, displaying a small thermal hysteresis and small shear during transformation. Thermodynamically, the free energy of the thermoelastic martensite transition is counterbalanced by strain energy in the parent phase, so that no plastic flow occurs during the phase transition but considerable stress develops.[3] This property permits the recovery of strain, associated with the growth of favorably oriented twinning variants in the lower symmetry martensite structure, when heated to the higher symmetry parent phase structure with a concomitant recovery of mechanical energy as well.

Among the thermoelastic martensites that display the SME, B2 intermetallics are the most prevalent. Two well-studied examples of B2 intermetallics are AuCd and NiTi, which have shape-memory applications. The B2 intermetallics (CsCl type structure with Pm$\underline{3}$m symmetry) are advantageous for studying the SME owing to their ordered structure.  In particular, the generation of dislocations (unrecoverable strain elements) and phase stability (retention of order because of sluggish diffusion kinetics) are two key components of the physical properties that make these alloys attractive.

Modifications to the microstructure are important to further optimize the SME in B2 intermetallics.  For example, pre-aging, cold-working, increased vacancy content, and decreased grain size are methods to optimize specific aspects of shape-memory behavior, illustrating the critical role of microstructure in the martensitic phase transition.[4] However, changes in microstructure are also known to affect the damping capacity, thereby altering the total internal friction and energy recovery associated with the martensitic transformation. [4]

The B2 alloys that undergo martensitic phase transitions display a large elastic shear modulus anisotropy, $A = {c_{44}}/{c^*}$ (for cubic symmetry), which is promoted by the weak resistance to c* (c* = 1/2 ($c_{11}$- $c_{12}$ ) ) shears. This type of phase transition can be interpreted through Zener's connection of a large $A$ value to a lattice instability.[5]  The B2 intermetallics that undergo a martensitic transition all display an anomalous temperature coefficient for the soft shear modulus, c*, indicating that even at several hundred degrees above the transition the structure has little resistance to deformations associated with c*. This resistance to deformation often decreases with temperature in a continuous fashion. The microstructural form of phonon softening is the tweed microstructures, and can serve as "frozen-in" embryos for martensite nucleation.[6]  Defects in the structure can interact with embryos to further reduce the nucleation barrier associated with the phase transition.[7]

Thermal effects play an important role in the nucleation of thermoelastic martensites. Therefore, a study of a shape-memory alloy at very low temperatures could help delineate fundamental features of the martensitic phase transformation. For example, at very low temperature, the thermal activation between the parent and product phase becomes small, and the effect of high-energy sites (i.e., grain boundaries and dislocations) are better isolated. The lowest proposed martensitic transition in a B2 intermetallic is that of AuZn.

The gold-zinc system displays a congruent melting point ($T_{melting}$ = 1058 K) near equal atomic composition, which permits single crystals of the ordered B2 phase to be readily prepared. At lower temperatures, the intermetallic reportedly displays a compositionally dependent phase transformation from 64 K (at compositions that are low in Zn) to 84 K (at Zn compositions greater than stoichiometry).[8] Although there is little published work on the fundamental thermodynamic properties and crystallographic features of the phase transformation, the reaction has been described as a martensitic transformation to a trigonal (P3) phase.[9] The linear, positive temperature coefficient of the c* elastic modulus from low-temperature to room temperature in the AuZn phase suggests an impending thermoelastic martensite transformation.[10] The 300 K equilibrium phase of AuZn is the B2 phase, unlike NiTi which must be quenched to retain the B2 phase, possibly locking defects and strains into the lattice. The ambient temperature B2 phase stability allows highly ordered, strain free single crystal samples of AuZn to be produced. The low temperature at which the transition occurs (~ 1/3 $Q_D$) reduces thermal energies and the possible effects of diffusion.

Resonant ultrasound spectroscopy (RUS) measurements were made on both single-crystal and polycrystalline samples to monitor the modulus and internal friction through the martensitic phase transformation. Mechanical testing at cryogenic temperatures was used to search for shape-memory effect in this alloy. Specific-heat measurements of the single-crystal RUS sample were made using a low-temperature semi-adiabatic calorimeter in order to investigate thermodynamic properties fundamental to the transformation in AuZn.

## II. EXPERIMENT

### A. Samples

Elemental Au and Zn metals in equal atomic composition, having purity of 99.9 at. % and 99.99 at. % respectively, were sealed at less than 1 Pa in a quartz tube. The inner quartz diameter was less than 1 mm in the sealed spiral section. After heating to 1373 K to melt the AuZn phase, the quartz spiral end was submerged in water for cooling. Directional solidification through the spiral extinguished all but one grain, permitting the bulk of the sample to solidify as a single crystal, from which parallelepiped samples ([100] x [010] x [001]) were cut. The same sample was used in both the elasticity and specific-heat measurements, and the weight of the parallelepiped sample was 96.5 mg. Polycrystalline samples were prepared in the same fashion, only the spiral turns were eliminated and the quartz tube (containing the molten alloy) completely submerged into a water bath.

## B. Resonant Ultrasound Spectroscopy

RUS is a technique in which the complete elastic tensor is determined from the spectrum of the mechanical resonances of an object. [11] A numerical procedure[12] is used to fit the expected frequency spectrum for a given set of elastic moduli, $c_{ij}$, and comparison with data of known crystal symmetries and orientations provides the matrix $c_{ij}$ consistent with the measurement. The cubic AuZn alloy has three independent moduli, which we take to be $c_{11}$, $c_{44}$ and $c^*$. These moduli are associated with the speeds of longitudinal and shear waves propagating along $c_{11}$, and shear waves along $c_{44}$ with displacements in $c^*$, respectively. Contact between the sample and the transducer is located at the corners of the parallelepiped. Thus the surfaces are unconstrained, so a single measurement may sweep through a martensitic phase transition despite surface relief developing at the transition. Peak widths provide internal friction data free from any scattering corrections.

## C. Mechanical Testing

It was anticipated that the AuZn alloy displayed elastic, crystallographic and thermodynamic features consistent with other reported B2 shape-memory alloys. Therefore, a simple test was performed to evaluate whether in fact this alloy had reversible plastic strain when deformed below the transformation temperature and re-heated. Compressive stress-strain tests using an Instron 1125 tester at $^4$He temperatures were used to verify whether AuZn is indeed a shape memory alloy. Cylindrical samples were compressed and unloaded below the transition temperature, and then remeasured after heating to room temperature to check for strain recovery. The strain rates used in the study were $10^{-3}$ s$^{-1}$.

**D. Specific Heat**

The specific heat was measured in zero magnetic field using an apparatus designed around a $^3$He insert capable of attaining temperatures as low as 0.4 K.[13] The sample was thermally attached to the sample platform with a thin layer of Apiezon N grease. The specific heat of the empty sample platform and the specific heat of Apiezon N [14] was measured separately and subtracted from the total specific heat in order to obtain the specific heat of the sample. The performance of the calorimeter was tested by measuring a single crystal of copper metal with a chemical purity of 99.99999 at.%. Below 30 K, the copper results were compared to zero-field low temperature data of Holste *et. al.* [15], Martin [16], and Phillips [17], and above 30 K the results were compared to the high-temperature measurements of Martin.[18] Our results fell within ±0.25% of the scatter obtained from the reference measurements, and our resolution was better than 0.1%. We estimate that the uncertainty in our measurements increased to ±4 % below 10 K, and ±1% in the transition region because of the small sample size. The resolution increases to 1% and 0.4%, respectively. A semi-adiabatic pulse technique was used to measure the specific heat from the lowest temperature up to 10 K, and to map the martensitic transition from $T_1$= 45 K to $T_2$= 70 K. An isothermal technique was used in the temperature region 10 K to 100 K.

# III. RESULTS AND DISCUSSION

## A. Elastic Properties

A plot of the temperature dependence of single crystal sample resonance frequencies is shown in FIG. 1(a). The frequencies of these modes are largely determined by the $c^*$ shear modulus, as it is the smallest elastic constant in the system by nearly an order of magnitude. The transition at 64 K occurs without significant precursor changes when approaching the transition temperature $T_{tr}$ from above (e.g. a soft mode). This result contrasts with measurements made on a large-grain size polycrystal sample, FIG. 1(b), where curvature appears as the transition is approached. This result may be attributed to strains generated by stresses evolved at the grain boundaries modifying the free energy near the transition. After cycling from 300 K to 4 K and back to 300 K, the resonances return to the same frequencies, indicating that the entire sample returns to an identical elastic state after cycling through the transition. We have determined the single-crystal elastic shear moduli for the temperature range above the transition. A linear fit to the moduli and anisotropy yields:

$c_{44}$ = 66.6 (± 0.2) - 0.037 (± 0.001)$T$ GPa ,

$c^*$ = 6.66 (± 0.01) + 0.0045 (± 0.0001)$T$ GPa ; and

$A$ = 10.0 (± 0.03) - 0.0097 (± 0.0002)$T$.

The data shown in FIG. 2 shows the cubic shear elastic moduli and also the shear anisotropy $A$, which varies from 7 at room temperature to 10 near the transition. The soft shear modulus, $c^*$, displays an anomalous positive temperature dependence. That agrees well with published pulse-echo ultrasound measurements.[19]

While our measured RUS peaks represent modes dominated by the shear moduli, the small longitudinal component in several modes provides a value for $c_{11}$, albeit with large errors. Attempting to fit $c_{11}$ and $c_{12}$ produces values some 10 % lower than the pulse echo data and also a positive temperature coefficient for $c_{11}$. Nonetheless, these data provide a bulk modulus that is in favorable agreement with a recent pressure-cell neutron scattering value of 122 GPa.[20]

The width of a resonance peak is proportional to the internal friction or dissipation in the sample. Because the sample is effectively isolated, this represents vibrational energy converted either to heat or transferred to other frequencies by nonlinear processes. Shown in FIG. 3 are details of the low-temperature behavior of the peak frequencies and typical widths in the polycrystal. Above the transition temperature there is very little dissipation and there is no maximum at the transition as might be expected, but a maximum in the temperature region 42 K to 60 K occurs. In this region mobile twin boundaries in the martensite may shift, absorbing the energy of the ultrasonic strains. Below 10 K the peaks are again sharp but we were unable to obtain consistent elastic fit to any symmetry.

Failure to fit suggests that the sample is inhomogeneous through its volume on a length scale comparable with the wavelength of the acoustic standing wave. The behavior below the transition does tell us that the modulus has approximately doubled, and there are no anomalous temperature coefficients. The transition occurs at $64 \pm 1$K with $\pm 1$K of thermal hysteresis. An expression for the long-wavelength $Q_D$ involving only the shear moduli [21] allows us to calculate a $Q_D = 207$ K from the $T = 0$ K linear extrapolation of the elastic moduli.

## B. Shape-Memory Effect at Low-Temperature

The compressive stress-strain tests performed using an Instron 1125 tester at $^4$He temperatures confirm that AuZn is indeed a shape memory alloy. Cylindrical samples were compressed and unloaded below the transition temperature to retain a plastic deformation. Length measurements on the samples were made after heating to room temperature and the low-temperature deformation. The stress-strain curves of polycrystalline AuZn performed at approximately 10 K (sample 1) and 50 K (sample 2) are shown in FIG. 4.. The sample exhibiting a higher flow stress was tested at $T\sim10K$, whereas the other sample was tested at $T\sim50K$. The initial yielding at both temperatures constitutes a plateau region up to 12% and 15% true strain. Such a region is a common feature in shape-memory alloys. The initial yield plateau is associated with the alignment of twins in the lower symmetry martensitic structure. The reversion of the aligned twinned structure upon heating to the higher symmetry phase gives rise to the shape recovery.

## C. Specific Heat

The specific heat measured in zero magnetic field of the AuZn martensitic transition is shown in FIG. 5. In this figure, the lattice specific heat has been subtracted from the total specific heat in order to clearly show the shape of the transition. The transition at 64.75 K is clearly a continuous transition with a characteristic lambda-shape similar to the ordering transition that occurs in $\boldsymbol{b}$-brass, another B2 alloy.[22] The electronic term and the $\boldsymbol{Q}_D$ were obtained from a fit to the low-temperature expansion of the Debye function given in (1):

$$C_V = \gamma T + B_3 T^3 + B_5 T^5 \qquad (1)$$

where the needed number of adjustable parameters for the lattice ($B_3$, $B_5$, ....., $B_n$) increases at high temperatures. The electronic and lattice specific heat was fitted from T=0.6K to $T$=10 K. A $\boldsymbol{Q}_D$ of 219 K, and $\boldsymbol{g}$ value of 0.38 mJ $K^{-2}mol^{-1}$ were obtained from this fit. In order to estimate the lattice specific heat in the transition region, $\boldsymbol{Q}_D$ and $\boldsymbol{g}$, obtained from the fit to (1), were substituted into a Debye-Einstein function [23],

$$C_p = (\ 3R\ [m\ D(\boldsymbol{Q}_D\,/\,T) + n\ E(\boldsymbol{Q}_E\,/\,T)] + \gamma T\ ) \tag{2}$$

where $m$, $n$, and $\boldsymbol{Q}_E$ are adjustable parameters; whereas $\boldsymbol{Q}_D$ and $\gamma$ were fixed at the values obtained from the low-temperature fit to (1). The fit of (2) to the specific-heat data was within experimental error, and it is shown in the inset of FIG. 5. The fit to the lattice was interpolated through the transition, and integration of $C/T$ from $T_1 = 47.5$ K to $T_2 = 67.0$ K gives an entropy of transition, $\Delta S_{tr}$, value of 2.02 J mol$^{-1}$ K$^{-2}$. A change in $\boldsymbol{Q}_D$ between the two phases becomes negligible, to the point that a single $\boldsymbol{Q}_D$ can be used to fit the lattice through the transition, shown in the FIG. 5 Inset.

## C. Geometric representation of the martensitic phase transition

Historically, thermoelastic martensitic transformations have been described as first-order displace mechanism, based on nucleation and growth mechanisms and discontinuous changes in the volume and shear elastic moduli through the transformations. However the lambda-like shape of the phase transformation in the specific heat data shown in FIG. 5 raises the possibility that the reduced temperature at which AuZn transforms makes it fundamentally different from other thermoelastic martensite. It appears that over a narrow Zn composition range the character of the martensitic transition lies at the borderline between that of a first-order (discontinuous) and a second-order (continuous) phase transition. Such an effect is certainly possible if the temperature is low because the transition is primarily shear-driven. In such a shear mechanism, two phases connected by a second-order phase boundary exhibit a continuous deformation from one phase to another by a Bain distortion.[24,25]

The composition dependence of the transformation temperature on heating ($A_s$, $A_F$) and cooling ($M_s$, $M_F$) based on the data of Pops and Massalski are shown in FIG. 6.[26]

It can be seen that the $M_s$ and $M_F$ curves are approximately 10 K apart at a mole fraction, $x = 0.56$ Zn. At $x = 0.49$ Zn, the $M_s$ and $M_F$ curves are separated by approximately 5 K. Finally at $x = 0.50$ Zn, the $M_s$ and $M_F$ curves are virtually on top of one another, at least within experimental error. Similarly, the heating curves ($A_s$, $A_F$) at $x = 0.56$ are separated by approximately 10 K, and at $x = 0.49$ and $x = 0.50$ Zn the heating curves are separated by approximately 5 K. Thus it appears that there is a very narrow separation in heating and cooling transformation temperatures at $x = 0.50$ Zn. There is no data reported in the region spanning equiatomic composition.

A schematic illustration based on the data of Ridley and Pops showing the variation of resistance with temperature for different Zn compositions spanning the equiatomic region are shown in FIG. 6 Inset (a).[8] A main feature observed in the resistivity is a continuous change in the martensitic transition at $x = 0.50$ Zn. Conversely at compositions less than $x = 0.50$ Zn , there was a smooth bump in the resistivity at $x = 0.497$ Zn and a sharp peak at $x = 0.495$ Zn. The same features are observed for compositions slightly greater than equiatomic Zn. Observation of the transformation in the equiatomic composition by an optical cold-stage microscope showed that the martensite platelets always retained a high degree of perfection.[25] This observation is consistent with de Haas-van Alphen oscillations.[27] In this study, Beck *et. al.* obtained good oscillations from an equiatomic alloy but not from alloys containing $x = 0.49$ or $x = 0.51$ Zn. Such oscillations should only be found when the lattice has a perfect site occupation. Thus it appears that the combination of equiatomic composition (removing a thermodynamic "bootstrap" from internal strains) and a low transformation temperature (decreasing both diffusion and entropy contributions) constrain the chemical potential and its derivatives to exhibit behavior intermediate between that of a first-order and second-order phase transition.

For these reasons we propose a two-dimensional phase space illustration for AuZn in Inset (b). In Au-Zn mixtures, the transformation temperature increases with increasing mole fraction $x$ of Zn, and below a temperature $T^*$, where a critical point occurs such that two first-order (discontinuous) curves meet. Above $x^*$ the transformation shows a slight decrease. Thus it appears that the two dashed lines are first-order transition lines that terminate in a critical region corresponding to $x^*$, $T^*$. Here, the equiatomic composition removes a mechanical shear driving term associated with inhomogeneity and the low-temperature removes the entropy drive. Similar types of phase behavior, which are grouped under the general heading *multicritical phenomena*, have been observed in ferroelectric materials[28] and in the simple metamagnet $FeCl_2$.[29]

## IV. CONCLUSION

AuZn provides a unique system for studying the martensitic phase transformation. The diminished influence of defects and thermal energy from the easy crystal growth, and low transition temperature leave a transition that has very little first-order character and displays a clear shape-memory effect, indicating that the stronger first-order features in other B2 martensites such as AuCd, and NiTi may be more a function of internal strains and the diffusive effects of thermal energy. The positive temperature coefficient of $c^*$ in the high temperature state has been remarked upon by Zener[5] as indicating an instability in the lattice. However, the unusually small elastic modulus $c^*$ provides a high phonon entropy for that transverse acoustic branch, lowering the free energy in the high temperature phase. Zener has suggested that this stabilizes the high temperature phase with diminishing effect at lower temperatures as the $TDS$ term in the free energy diminishes, consistent with our measurements.[20] We note that the positive gradient exists at temperatures hundreds of degrees above the transition, and that in the single-crystal case the behavior is linear and there is no indication of the impending transition. The Debye temperature is in favorable agreement with the value of 217 K for AgZn, although the electronic specific heat for AuZn is nearly half of that reported for AgZn. [30]

## ACKNOWLEDGEMENT

The authors would like to thank Dr. Loren Jacobson, Dr. Robert Field, Prof. Harold Stokes, Dr. Angus Lawson, and Dr. James L. Smith for informative discussions. We thank Mr. Larry Hults for sample preparation and measurement assistance. JCL thanks Professor N. E. Ridley, Professor T. B. Massalski, and TMS for permission to reprint the data shown in FIG. 6. This work was carried out at Brigham Young University and also at the Los Alamos National Laboratory. All work was performed under the auspices of the United States Department of Energy.

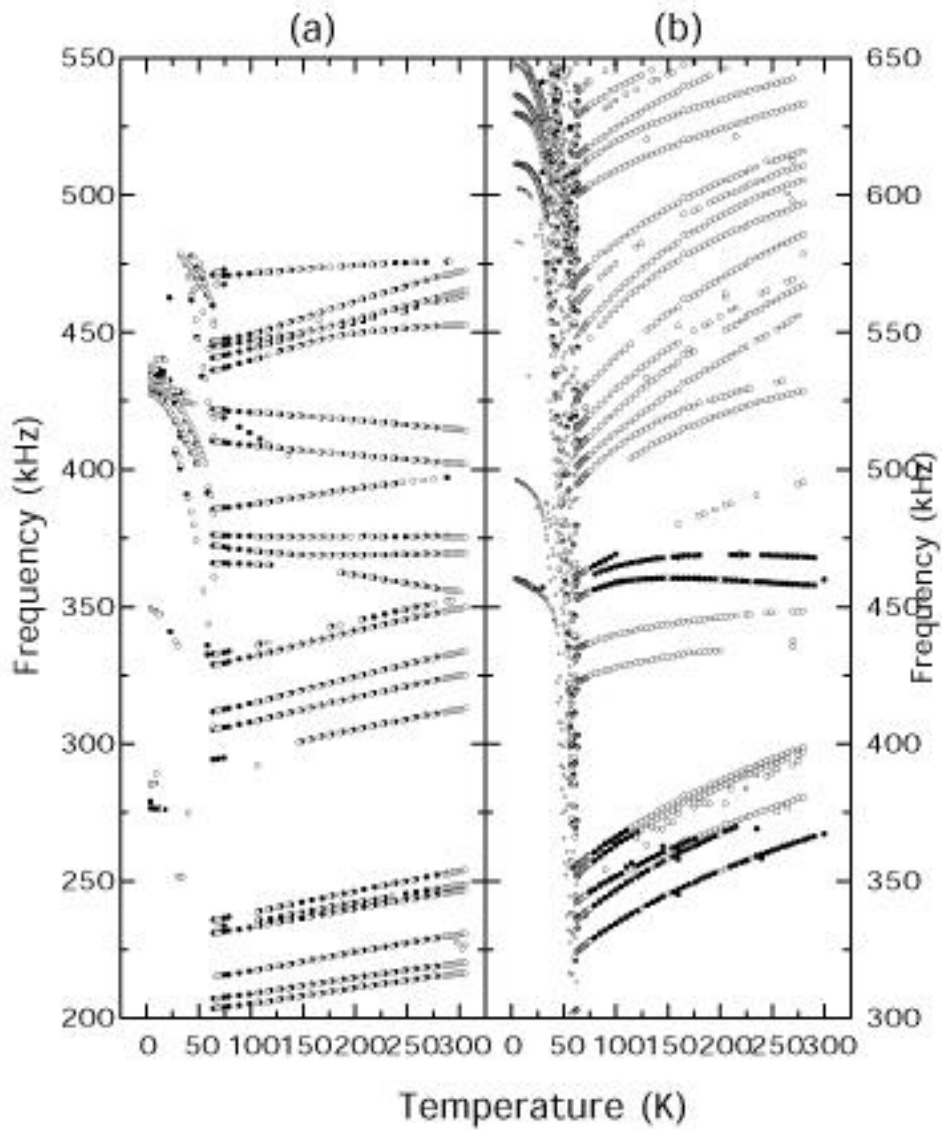

FIG. 1.  RUS resonance frequencies for (a) single crystal and (b) polycrystal AuZn as a function of temperature. Positive gradients are associated with modes dominated by c*, negative by $c_{44}$.

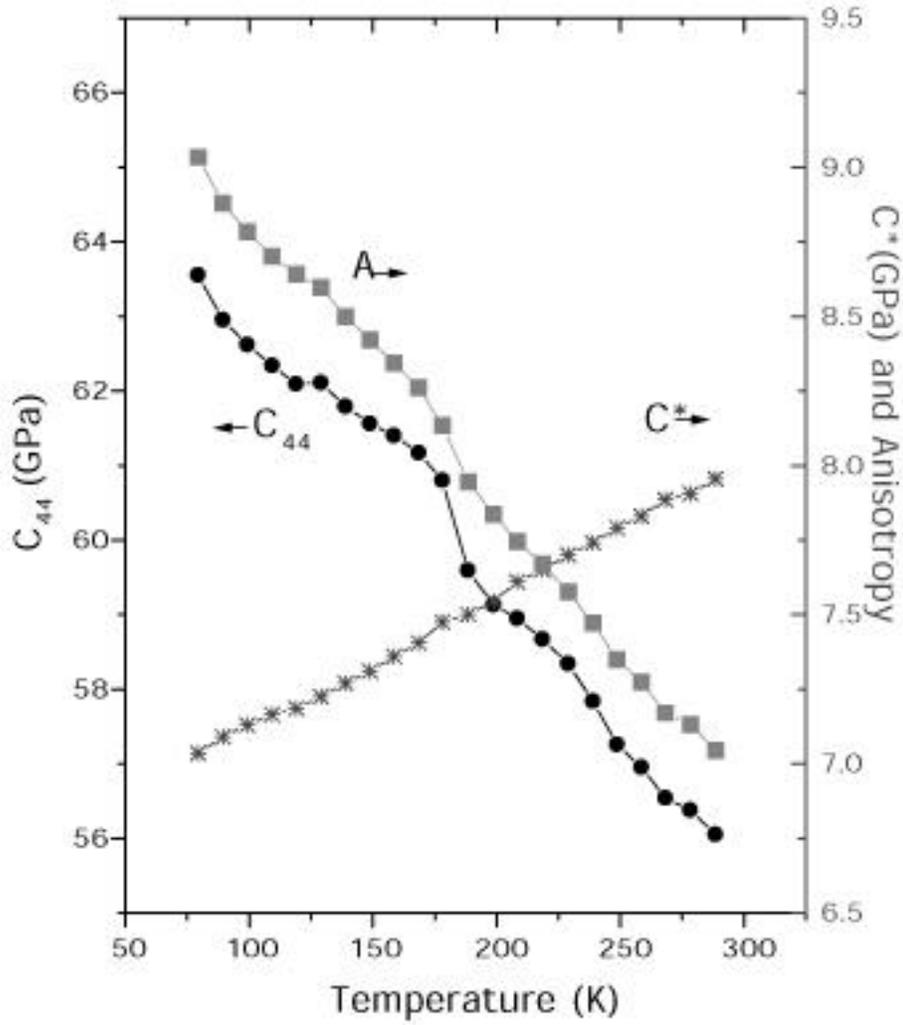

FIG. 2. The data fitted to the shear moduli and the shear anisotropy. One can see that the soft shear modulus C* displays an anomalous positive temperature dependence.

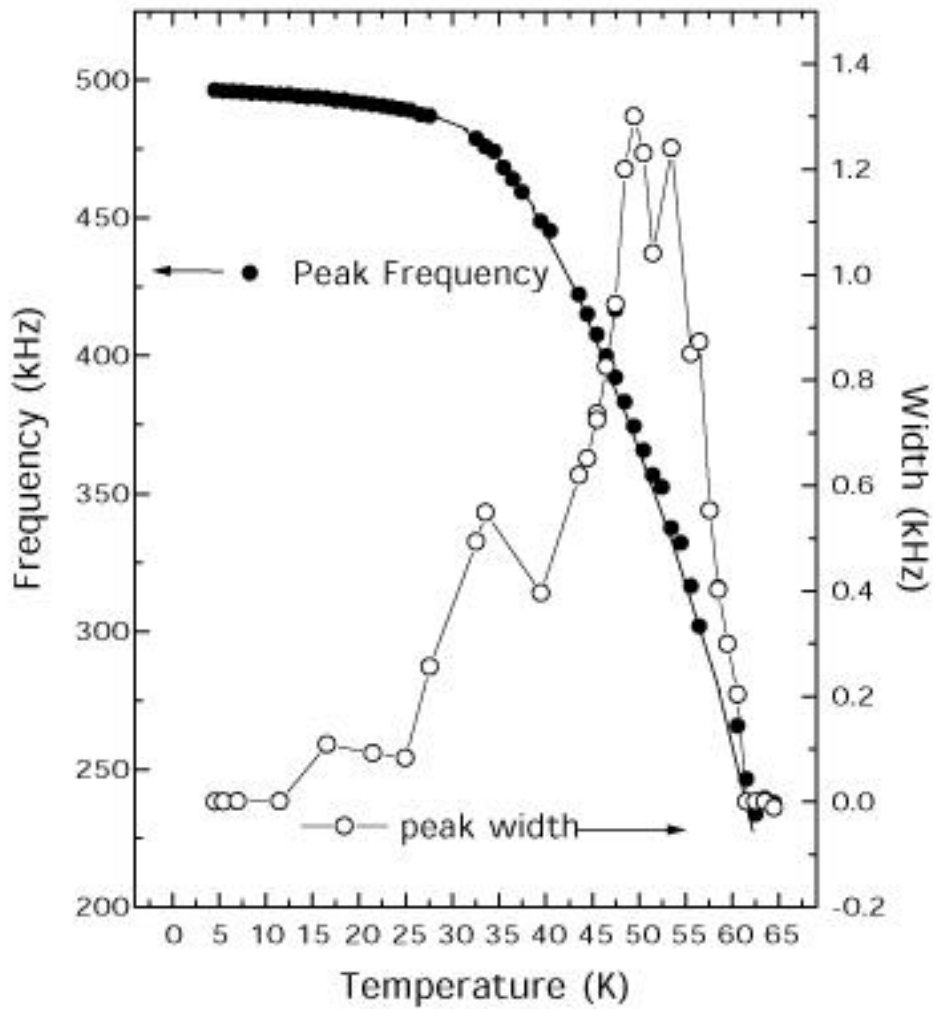

FIG. 3. RUS resonance frequencies and typical peak width (internal friction) below $T_{\text{tr}}$. The low value of the dissipation continues from below $T_{\text{tr}}$ to 300K. There is no peak at the transition temperature.

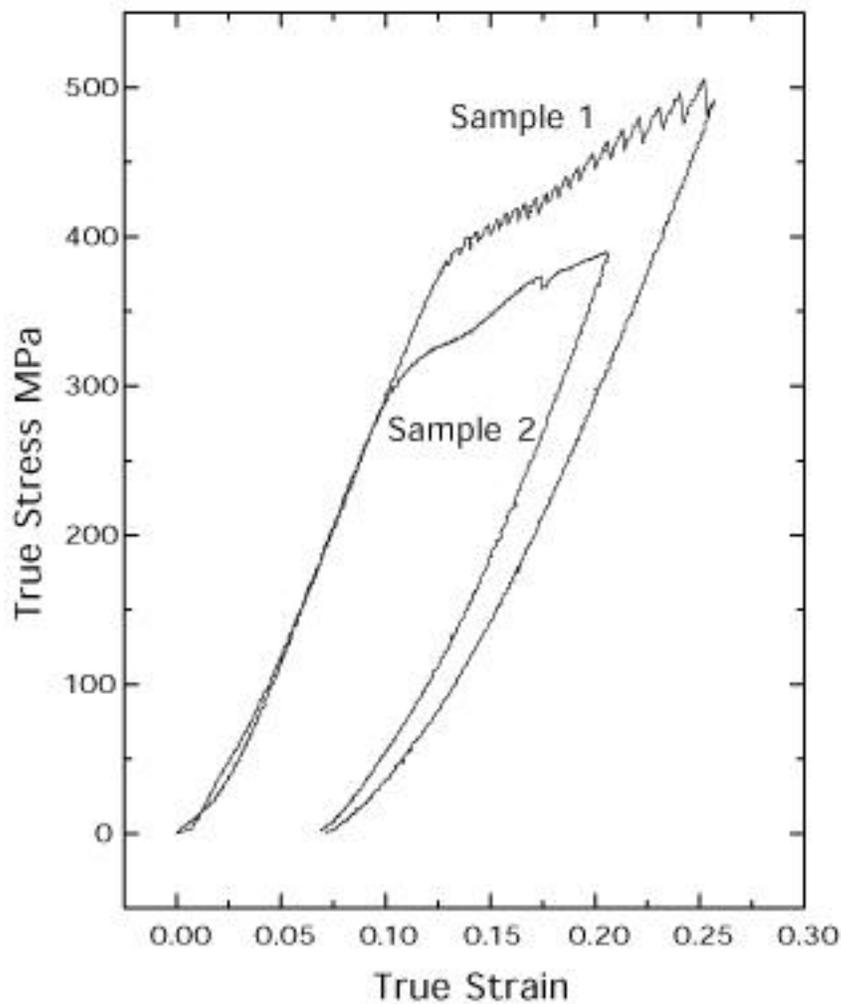

FIG. 4. Stress-strain cycles at constant temperature for two polycrystal AuZn samples showing approximately 7 % retained true strain. The axes are offset to more clearly illustrate the loop. Sample 1 was compressed to ~12 % true plastic strain and unloaded at $T \sim 10$ K. Sample 2 was compressed to ~10 % true plastic strain and unloaded at $T \sim 50$ K. The length of both samples was measured after warming above the phase

transformation temperature, showing that approximately half of the retained plastic strain was recovered.

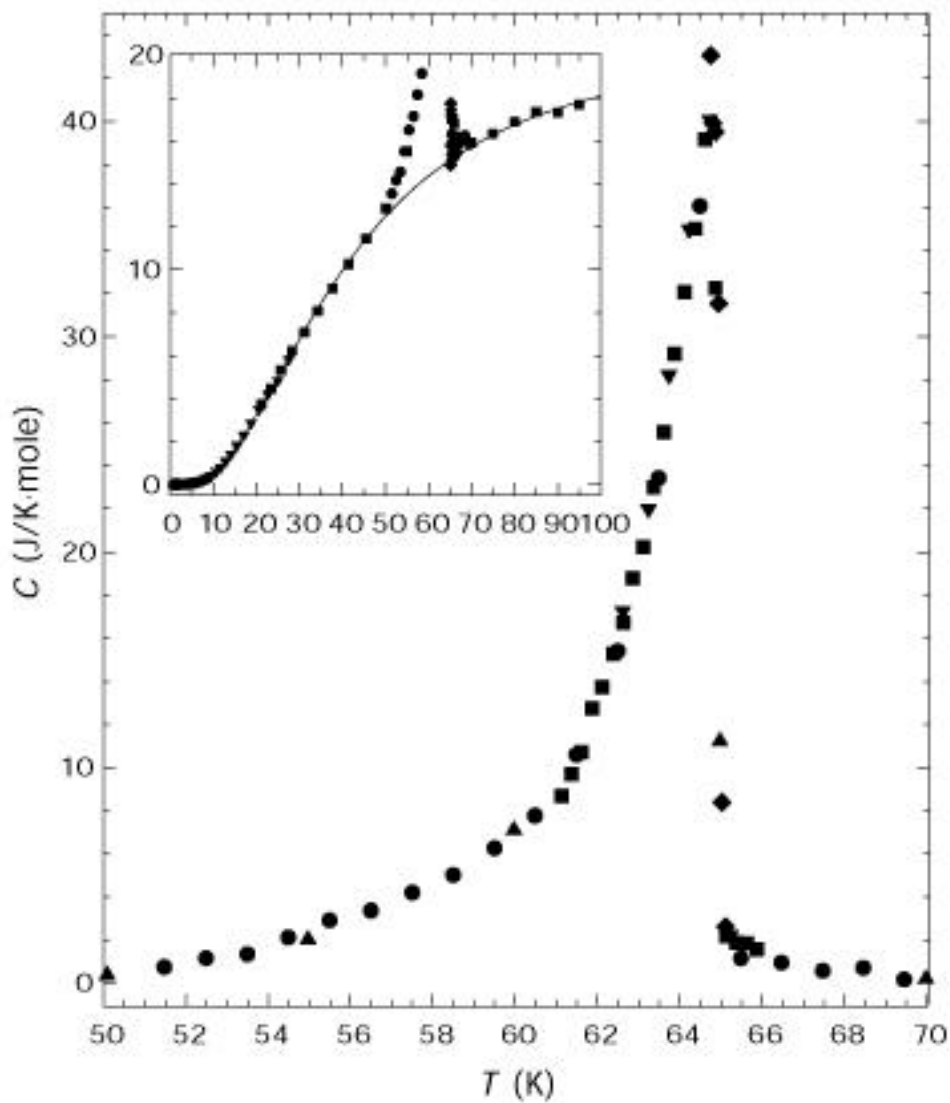

FIG. 5. The specific heat of the martensitic transition is shown with the lattice contribution subtracted. There were six runs that were made through the transition, each using a smaller $\Delta T$, denoted by different symbols. Between each run, the sample was cooled at a rate of 5 K hour$^{-1}$ to ensure reproducibility in the specific-heat data through the transition. For the martensitic transition at 64.75 K, a temperature hysteresis of 0.89 K ($\pm 0.025$ K) was measured. Integration of $C/T$ gives an entropy of transition, $\Delta S_{tr}$, value of 2.02 J mol$^{-1}$ K$^{-2}$ **Inset:** The fit of the lattice specific heat to the martensite and austenite phases. This fit was interpolated through the transition.

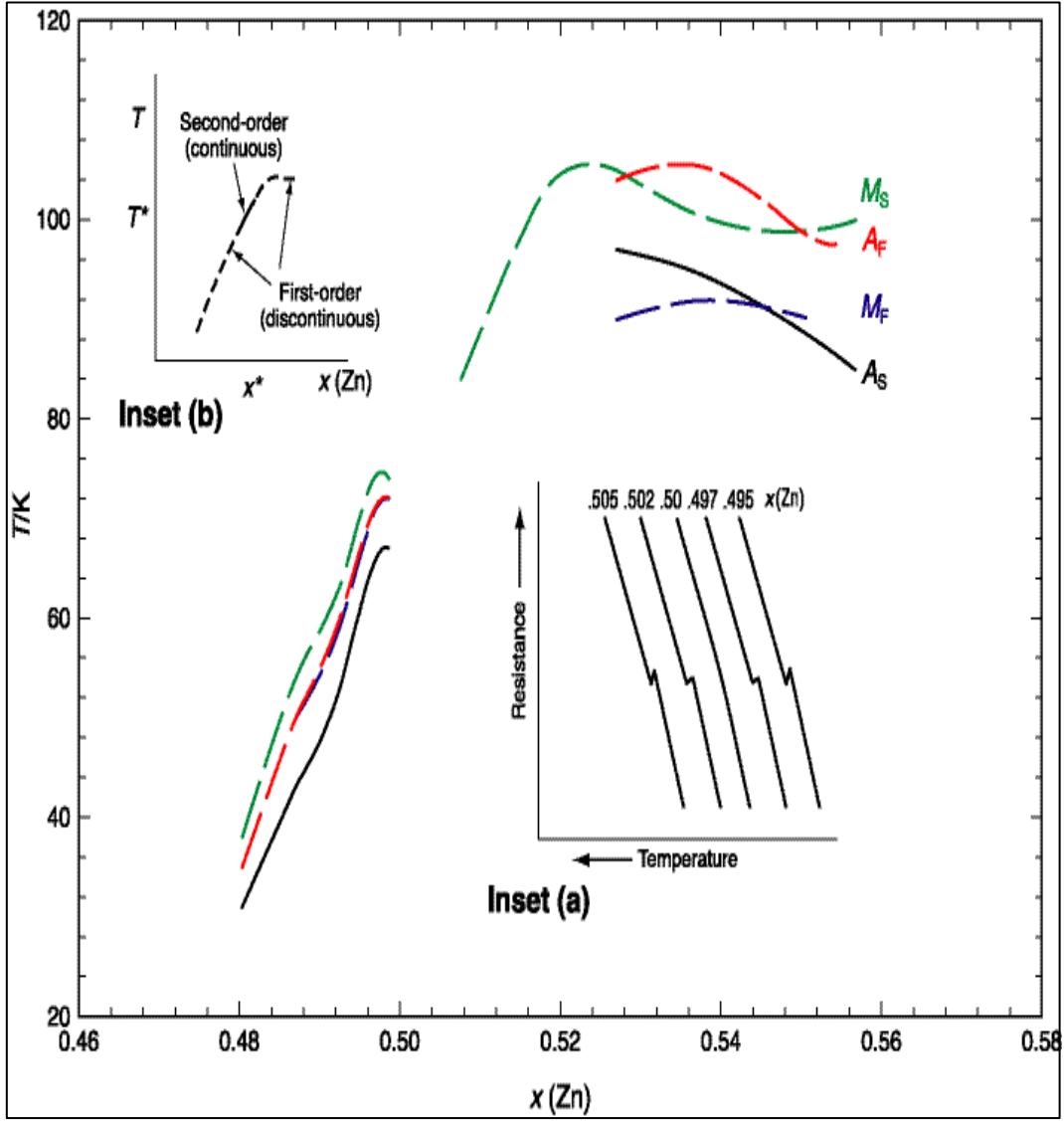

FIG. 6. The composition dependence of Zn to the transformation temperature on cooling and heating. We note that for near equiatomic composition there are no values for transformation temperatures (*figure reprinted by permission*, Horace Pops and T. B. Massalski, Trans. Met. Soc. AIME, 233, 728 (1965)). **Inset (a)**: Schematic illustration showing the variation in resistance with temperature for different Zn compositions (*figure reprinted by permission*, Norman Ridley and Horace Pops, Met. Trans.,1, 2867 (1970)).

**Inset (b)**: Temperature composition representation illustrating the nature of the composition dependent martensitic phase transformation showing two lines of first-order phase transitions terminating in a critical point at ($x$ *,$T$*). Here there is no distinction between transformation temperatures on heating and cooling.

---

[1] K. P. Chong, S. C. Liu, and O. W. Dillon, US-Japan Workshop on Smart Materials and Structures, University of Washington, 3, (1995), edited by Kanryu Inoue, S. Shen, and M. Taya, TMS, Warrendale, (1996).

[2] C. M. Wayman and H. K. Bhadeshia, in *Physical Metallurgy*, **Fourth** edition, edited by Robert W. Cahn and Peter Haansen' (Elsevier Science, Amsterdam, 1996), Vol. **2**, Chap. 16, p.1508.

[3] J. Ortin, in *Mechanics of Solids with Phase Changes*, **First** edition, edited by M. Berveiller and F. D. Fischer (Springer, New York, 1997), Vol. **1**, Chap. 1, p.1.

[4] T. Tadaki, in *Shape Memory Alloys*, **Second** edition, edited by K. Otsuka and C. M. Wayman (Cambridge University Press, Cambridge, 1999), Vol. **1**, Chap. 4, p.97.

[5] C. Zener, Phys. Rev. **71**, 846 (1947).

[6] L. Kaufman and M. Cohen, Prog. in. Met. Phys. **7**, 165 (1958).

[7] P. C. Clapp, Phys. Stat. Sol. (b) **57**, 561 (1973).

[8] Norman Ridley and Horace Pops, Met. Trans. **1**, 2867 (1970).

[9] T. Makita, A. Nagasawa, Y. Morii, N. Minakawa, and H. Ohno, Phys. B. **213 & 214**, 430 (1995).

[10] T. B. Massalski, Acta Met. **15**, 1770 (1967).

[11] A. Migliori and J. L. Sarrao, *Resonant Ultrasound Spectroscopy* (Wiley, New York, 1997)

[12] W. M. Visscher, A. Migliori, T. M. Bell, and R. A. Reinert, J. Accoust. Soc. Am. **90**, 2154 (1991).

[13] J. C. Lashley, Mechanics of metals with phase changes, thesis, Brigham Young University, *LA-13788-T*, 19-25, (http://www.doe.gov/bridge) (2000).

[14] M. Widn and N. E. Phillips, Cryogenics **15**, 36 (1975).

[15] J. C. Holste, T. C. Cetas, and C. A. Swenson, Rev. Sci. Instrum. **43**, 670 (1972).

[16] D. L. Martin, Phys. Rev. B. **8**, 5357 (1973).

[17] N. E. Phillips, J. P. Emerson, R. A. Fisher, J.E. Gordon, B. F. Woodfield, and D. A. Wright, *unpublished data.*

[18] D. L. Martin, Rev. Sci. Instrum. **58**, 639 (1987).

[19] R. J. Schiltz, T. S. Prevender, and J. F. Smith, J. Appl. Phys. **42**, 4680 (1971).

[20] R. D. Field, private communication.

[21] H. Siethoff, Intermetallics **5**, 625 (1997).

[22] M. B. Salamon and F. L. Lederman, Phys. Rev. B. **10**, 4492 (1974).

[23] B. F. Woodfield, J. L. Shapiro, R. Stevens, J. Boerio-Goates, R. L. Putnam, K. B. Helean, and A. Navrotsky, J. Chem. Thermodyn. **31**, 1573 (1999).

[24] Edgar C. Bain, *Pioneering in Steel Research: A Personal Record*, **1**st edition (American Society Metals, Metal Park, 1975).

[25] Albert Migliori, Joseph p. Baiardo, and Timothy W. Darling, Los Alamos Science: Challanges in Plutonium Science **1 (26)**, 208 (2000).

[26] Horace Pops and T. B. Massalski, Trans. AIME **233**, 728 (1965).

[27] A. Beck, J. P. Jan, W. B. Pearson, and I. M. Templeton, Phil. Mag. **8**, 351 (1963).

[28] George A. Rossetti, Jr. and Alexandra Navrotsky, J. Sol. Stat. Chem. **144**, 188 (1999).

[29] J. Skalyo, Jr., A. F. Cohen, S. A. Friedberg, and R. B. Griffiths, Phys. Rev. **164**, 705 (1967).

[30] J. P. Abriata, O. J. Bressan, C. A. Luengo, and D. Thoulouze, Phys. Rev. B. **2**, 1464 (1970).